\documentclass[pdflatex,sn-chicago]{sn-jnl}

\jyear{2021}%

\theoremstyle{thmstyleone}%
\usepackage{lettrine}
\theoremstyle{thmstyletwo}%

\theoremstyle{thmstylethree}%
\DeclareMathAlphabet{\mathbfsf}{\encodingdefault}{\sfdefault}{bx}{n}

\begin{document}

\title[\it Gradient of a Vector Field]{\it A Note on the Gibbsian Representation of the Gradient of a Vector Field}


\author*[1]{\fnm{\textsc{Brian}} \sur{\textsc{Wood}}}\email{brian.wood@oregonstate.edu}

\author[2]{\fnm{\textsc{Peeter}} \sur{\textsc{Joot,}}}

\author[3]{\fnm{\textsc{Stephen}} \sur{\textsc{Whitaker}}}


\affil[1]{\orgdiv{School of Chemical, Biological, and Environmental Engineering}, \orgname{Oregon State University}, \orgaddress{ \city{Corvallis}, \postcode{97331}, \state{OR}, \country{USA},\\ \city{  https://orcid.org/0000-0003-3152-7852 }}}

\affil[2]{\orgaddress{\city{Toronto}, \postcode{M5S 1A1}, \state{Ontario}, \country{Canada}}}

\affil[3]{\orgdiv{Department of Chemical Engineering (emeritus)}, \orgname{University of California, Davis }, \orgaddress{ \city{Davis}, \postcode{95616}, \state{CA}, \country{USA}}}



\abstract{In this note, we provide a important considerations of a familiar topic: the gradient of a vector field.  The gradient of a vector field is a common quantity represented in continuum mechanics. However, even for Cartesian coordinate systems, there are two different representations for this quantity in common use, which leads to ambiguity in some results.  We review the history leading to the \emph{Gibbsian representation} for the gradient of a vector field, and provide some suggestions to help clarify the meaning of such terms when represented in conventional Gibbsian vector or tensor notation.  In an appendix, we briefly expand on the connection with the Gibbsian representation of the deformation and rotation tensors in the framework of geometric algebra (GA). }


\maketitle
\setcounter{section}{1}

\lettrine[findent=-2pt, lines=3,lhang=-0.33]{\bf T}~he contributions to both physics and mathematics by J. William Gibbs is somewhat astounding in its creativity and breadth.  While Gibbs is perhaps most well known for his work on thermodynamics and the foundations of statistical mechanics, his contributions to classical vector analysis were both prescient and far-reaching.  One of Gibbss' primary interests in his developments was to generalize the theory to handle second-order Cartesian tensors.  This led to the notion of \emph{dyads} (and \emph{dyadics}, which are the sum of dyads), which represented a logical extension of vector algebra. Gibbs' motivation appears to have been to give concrete meaning to the following between operation among vectors ${\bf a}$, ${\bf b}$, and ${\bf r}$

\begin{equation}
    {\bf a} ({\bf b}\cdot\bf{r}) =  ({\bf a} {\bf b})\cdot\bf{r}
\end{equation}
where here the dot represents the conventional Gibbsian dot product between two vectors.  Gibbs then defines the new product ${\bf a} {\bf b}$ as follows \citet[][\textsection 99]{gibbs1901vector}.  This development is reported in detail both in his early treatise on vector algebra \citep[][Chp.~III]{gibbs1884vector}, and in the book produced by E. B. Wilson based on Gibbs' lectures on the topic at Yale University in 1899-1900 \citep[][Chp.~V]{gibbs1901vector}.  Gibbs provides the following definition for the dyadic product \citep[][p.~265]{gibbs1901vector}

\begin{quote}
    {\it Definition} An expression ${\bf a}{\bf b}$ formed by the juxtaposition of two vectors with out the intervention of a dot or cross is called a dyad...  The first vector in a dyad is called the {\it antecedent}; and the second vector, the {\it consequent}.
\end{quote}
With this definition in combination with his previously established rules for vector algebra, Gibbs was able to construct an algebra for second-order tensors in Cartesian space, and to provide a number of useful identities that are still used routinely in mechanics (e.g., the Stokes-Gibbs decomposition for second-order tensors discussed below; the definition of the set of basis tensors [nonions] for second-order tensors in Cartesian coordinates; see Appendix \ref{appendixA} for discussion of the latter development).

It is clear that Gibbs saw that his theory of dyadics could be extended to include higher-order tensors (he mentions \emph{triadics} and \emph{polyadics} in passing in \citet[][\textsection 104,p.~281]{gibbs1901vector}), he did not himself extend the theory substantially beyond dyadics (possibly because he had no immediate applications requiring him to do so).  This extension was, however, conducted with some rigor in a paper by F. Hitchcock in 1923 \citep{hitchcock1923}.

Although Gibbs' extension of  vector algebra to form a well-defined algebra of second-order tensors is interesting topic itself for historical reflection (see \citet{crowe1967history} for this history), the purpose of this paper is more modest and specific. 
Here, we provide some technical clarity on a relatively familiar topic: the gradient of a vector field.  While the topic is, seemingly, a relatively basic one, there is nonetheless need for additional discussion and perspective in the literature.  Gibbs himself left no confusion about the topic in his own writings.  However, as is sometimes the case, an alternate definition for the gradient of a vector field (which is a dyadic form) has been erroneously identified as the Gibbsian result.  The presence of competing definitions certainly causes confusion in the literature,  and discussion of Gibbs' original approach and definitions for the quantity  can  help to bring additional clarity to the issue.

In the remainder of this note, we discuss the definition of Gibbs' dyadics, and how Gibbs defined the gradient of a vector.  This, then, represents \emph{the canonical form for expressing the gradient of a vector in Gibbs notation}.  Because Gibbs himself was very interested in the extension of these ideas into what was called at the time \emph{multiple algebra} \citep{gibbs1886multiple}, we also provide a brief explanation about how Gibbs' dyadic notation can be rectified with the work done by Grassmann and Clifford, now generally referred to as geometric algebra (GA).   In the material following, when tensor components are explicitly listed, they are constructed for a Cartesian coordinate system only; primarily this is to keep the concepts clear and uncluttered by metric tensors.  The results, however, do not depend at all on the coordinate system employed.

\subsection{Gibbs' dyadic}

The definition of the dyadic of two vectors is rarely misunderstood in the modern literature; the definitions given above, as first outlined by \citet{gibbs1884vector} are usually those that are adopted in modern representations.  For specificity, we note explicitly that for two vectors ${\bf a}$  and {\bf b}, the dyad is formed by (cf. \citet[][\textsection 101]{gibbs1901vector})

\begin{align}
    {\bf a}{\bf b} &\equiv {\bf a}\otimes{\bf b} \nonumber \\
    &= a_i b_j {\bf e}_i\otimes {\bf e}_j\nonumber \\
     &=a_1b_1{\bf e}_1\otimes{\bf e}_1 + a_1b_2{\bf e}_1\otimes{\bf e}_2+a_1b_3{\bf e}_1\otimes{\bf e}_3 \nonumber\\
&+ a_2b_1{\bf e}_2\otimes{\bf e}_1 + a_2b_2{\bf e}_2\otimes{\bf e}_2+a_2b_3{\bf e}_2\otimes{\bf e}_3\nonumber\\
&+a_3b_1{\bf e}_3\otimes{\bf e}_1 + a_3b_2{\bf e}_3\otimes{\bf e}_2+a_3b_3{\bf e}_3\otimes{\bf e}_3
\label{dyadform1}
\end{align}
Here, ${\bf e}_i\otimes{\bf e}_j$ represent a basis for second-order tensors as described by Gibbs \citep[][Chp.~V, \textsection  101]{gibbs1901vector};  additional details about the Gibbs representation of basis dyads is given in Appendix \ref{appendixA}.
In conventional matrix representation, the dyadic (a second-order tensor) is given by
\begin{align}
{\bf a}\otimes{\bf b}
&=
    \begin{pmatrix}
    a_1 b_1 & a_1 b_2 & a_1 b_3 \\
     a_2 b_1 & a_2 b_2 & a_21 b_3 \\
     a_3 b_1 & a_3 b_2 & a_3 b_3 \\
    \end{pmatrix} 
\end{align}
Here, and in the material following, we have adopted the modern use of the symbol ``$\otimes$'' to explicitly  indicate the formation of Gibbsian dyad.  We have  also adopted  an index form for Gibbs' unit vectors ${\bf i} ={\bf e}_1$, ${\bf j} ={\bf e}_2$, and ${\bf k} ={\bf e}_3$.  While these notations were not adopted by Gibbs, they have become a regular convention in typesetting for vectors and dyads, and for mixing index notation with the notation of Gibbs.  The use of the ``$\otimes$" symbol has become more than superfluous notation with the increasing use of geometric algebra (GA), in which the product of two vectors with no intervening symbol defines the \emph{geometric product}; thus explicit use of the ``$\otimes$" symbol helps to eliminate potential confusion.  While Gibbs' terminology \emph{dyadic} is used here, this same operation is also known as the \emph{direct product} \citep{simmonds2012}, the \emph{tensor product} \citep{goodbody1982}, and occasionally the \emph{exterior product} \citep{degroot1963}.

Clearly, ${\bf a}\otimes{\bf b} $ is not, in general, a symmetric quantity.  Because of this lack of symmetry, multiplying a vector by the dyadic from the left and from the right, in general, yield different results.   Gibbs again addresses this at some length.  Defining $\mathbfsf{T}={\bf a}\otimes{\bf b}$, then one has a dyadic in \emph{postfactor} form defined by Gibbs when written as

\begin{align}
      {\bf c}  \cdot \mathbfsf{T}&  =  {\bf c}\cdot({\bf a} \otimes {\bf b})
\end{align}
Analogously, the dyadic is in \emph{prefactor} form if
\begin{align}
       \mathbfsf{T}\cdot {\bf c}& =({\bf a}\otimes {\bf b})\cdot {\bf c}  
\end{align}
This definition can be seen explicitly in the work of \citet[][ \textsection 101]{gibbs1901vector}.  This also serves to define left (prefactor) and right (postfactor) multiplication of a vector and a matrix.
Note that because these operators are in general non-commutative we have $ \mathbfsf{T}\cdot {\bf c}  \ne {\bf c}  \cdot \mathbfsf{T}$.  However, by defining the transpose  (conjugate) of the dyad \citep[][\textsection 100, p. 266]{gibbs1901vector}, one finds

\begin{align}
       {\bf c}\cdot\mathbfsf{T} &=  \mathbfsf{T}^\dagger\cdot {\bf c}
\end{align}

\subsection{The indeterminate nature of dyadics}

The prefactor and postfator forms above indicate that, in general, forming the dot product (contraction) of a dyadic and a vector is not a commutative operation.  A dyadic is then, in an operational sense, is not fully specified until it is also indicated \emph{how the dyadic is to be used} in subsequent operations.  Gibbs \citep{gibbs1901vector} states this clearly in  \textsection 102 (page 271)

\begin{quote}
    The symbolic product formed by the juxtaposition of two vectors ${\bf a}$, ${\bf b}$ without intervention of a dot or cross is called the \emph{indeterminate product} of the vectors ${\bf a}$ and ${\bf b}$.  The reason for the term indeterminate is this... the product ${\bf a b}$...acquires a determinate physical meaning only when used as an operator.
\end{quote}
The indeterminate nature of the dyadic product was an important facet of analysis via dyadics, and one that Gibbs  emphasized a number of times in his work.

\section{The gradient of a vector field in Gibbsian form}

The gradient (grad) of a continuous (or, at least, $c_1$ continuous) field $f(x,y,z)$ in a Cartesian coordinate system $(x,y,z)$ is defined by

\begin{equation}
    \mathrm{grad}(f)\equiv \left(\frac{\partial f}{\partial x}{\bf i}+\frac{\partial f}{\partial y}{\bf j}+\frac{\partial f}{\partial z}{\bf k} \right) 
\end{equation}
As a matter of notation, the \emph{nabla} operator, $\nabla$, seems to have been established in both the quaternion literature (by P.G. Tait \citeyear[][p.~221]{tait1867elementary}) and in the vector analysis literature (e.g., \citet{gibbs1884vector}).  Thus, one finds the compressed notation (e.g., \citet[][Chp.~IV]{gibbs1884vector})

\begin{equation}
\nabla f(x,y,z) \equiv \left(\frac{\partial f}{\partial x}{\bf i}+\frac{\partial f}{\partial y}{\bf j}+\frac{\partial f}{\partial z}{\bf k} \right) 
\end{equation}\\
and, subsequently, the definition of the gradient as a vector operator

\begin{equation}
\nabla \equiv \left(\frac{\partial }{\partial x}{\bf i}+\frac{\partial }{\partial y}{\bf j}+\frac{\partial }{\partial z}{\bf k} \right) 
\end{equation}\\

While the dyadic formed by two vectors is usually not misinterpreted, somewhat unexpectedly, when it comes to the matter of the dyadic of the \emph{gradient operator} and a \emph{vector} (in contrast to the dyadic of two vectors), the interpretation of the dyadic has split into two competing schemes.  Both the representation as suggested by Gibbs and \emph{the transpose of this representation} have been adopted in the literature with nearly equal frequency.  Adding to the confusion is the fact both of them are frequently referred to as being consistent with Gibbs vector notation.  

Thus, one finds that the symbol $\nabla \otimes {\bf v}$ is given the following definition [examples adopting this definition include \citet[][Chp. IV]{gibbs1884vector}, \citet[][p.~404, Eq.~(3)]{gibbs1901vector}, \citet[][p.~6, Eq.~(3.1)]{truesdell1954}, \citet[][Eq.~11.2]{serrin1959}, \citet[][p18]{chapman1961}, \citet[][p. 455]{degroot1963}, \citet[][p.~44]{milne1968}, \citet[][p.~109]{goodbody1982}, and  \citet[][p.~2]{pozrikidis2011}]

\begin{align}
  \hspace{-10mm}&\textrm{Gibbsian definition}& \hspace{-20mm} \nabla \otimes {\bf v} \rightarrow \frac{\partial v_j}{\partial x_i}{\bf e}_i \otimes {\bf e}_j
  \label{gibbsform}
\end{align}
and one also finds the following \emph{alternate definition} [examples of its use can be found in \citet[][cf. Eq.~(1.2.2) and index representation on p. 34]{batchelor1953}, \citet[][p.~121, Eq.~(6.3.3)]{aris1962vectors}, \citet[][p.~637, Eq.~(1-4)]{slattery1972}; \citet[][p.~384, Eq.~(17-2)]{mcquarrie1976}; \citealp[][p.~23, Eq.~(2.71)]{pope2000}; \citet[][p.~124]{lautrup2005}; \citet[][p.~50, Eq.~(2.28.4)]{lai2009}; and \citet[][p.~64, Eq.~(2.28.4)]{wegner2009}]

\begin{align}
   \hspace{-10mm} &\textrm{Alternate definition}& \hspace{-20mm}  (\nabla \otimes {\bf v})^* \rightarrow \frac{\partial v_i}{\partial x_j}{\bf e}_i\otimes {\bf e}_j
    \label{altform}
\end{align}
Occasionally, both definitions have appeared in the same text [\citealp[][p.~731, Eqs.~A.4-27 and A.4-28]{bsl1960}].  The superscript ``*" is not normally used in typesetting, and has been added here only to distinguish that these two quantities are different from one another.   

\subsection{Indeterminacy in the gradient of a vector}
It is understandable why this bifurcation in representations occurred; as stated above, the dyadic is inherently indeterminate, and one may erroneously come to associate the transpose (complement) of the gradient of the vector field with the gradient of the vector field itself.  Adding somewhat to the confusion,in modern representations of linear algebra matrices (whether dyadics or not) are conventionally in prefactor position, sometimes referred to as \emph{right multiplication}.  However, \emph{left multiplication} of a vector and a matrix (or dyadic) is also a well-defined operation (cf. \citet[][Chp.~III, \textsection 109]{gibbs1884vector}).  It is the assumption of left multiplication in the definitions provided by Gibbs that has created some of the confusion regarding the representation of the gradient of a vector field.  

The Gibbsian definition amounts to a 
\emph{choice} adopted to establish a formalism.  However, Gibbs did not make this choice arbitrarily.  For a vector field ${\bf v}$, his particular choice of definition for $\mathrm{grad}({\bf v})\equiv \nabla \otimes {\bf v} $ reflects the convention that an operator acts on object from the right. It is  not  by chance that this is operationally identical to that for the dyadic formed by two vectors, ${\bf a}\otimes {\bf b}$, where one treats the \emph{operator} $\nabla$ as though it represented the components of a conventional vector.  

The indeterminate nature of the dyadic formed by the gradient of a vector field was apparently well known to early proponents of vector analysis.  As an example  \citet[][]{weatherburn1920} (which might now seem an extreme measure) took special efforts to recognize this by defining the conjugate operators $\nabla \otimes {\bf v}$ and ${\bf v}\otimes \nabla$. 

The primary problem with these two competing definitions is not one of establishing \emph{which} form of the tensor is correct; rather, it is problem that requires one to recognize that the dyadic product, rendered by itself, is  \emph{indeterminate} in the sense discussed above.  
 
While apparently well understood in the past, the indeterminate property dyadics seemed to be partly lost from the modern understanding of vector algebra (as evidenced by the and competing definitions for the gradient of a vector field). At some point in the past, confusion arising  from the two competing  definitions for the gradient became deeply embedded in the literature.  In one well-respected text on continuum mechanics, the author underscores this uncertainty by explicitly stating \citep[][p.~637, footnote 1]{slattery1972}

\begin{quote}
    Unfortunately, while I believe this [Eq.~\eqref{altform} of this paper] to be the most common meaning for the symbol $\nabla {\bf v}$, some authors define \ldots [Eq.~\eqref{gibbsform} of this paper].  Where we would write $(\nabla{\bf v})\cdot{\bf w}$, they say instead ${\bf w}\cdot(\nabla{\bf v})$.
\end{quote}

While Gibbs' vector algebra has continued to be one of the primary means of compactly expressing complex equations in many areas of physics, mathematics, and engineering, the dyadic product remains a tool that appears somewhat infrequently (perhaps because formulations avoiding the dyadic can often be found).  The infrequency of its use has almost certainly added to the misunderstandings that appear in the literature when it is adopted.

\subsection{An Example: The Differential Rate of Strain}

One of the most familiar examples where the gradient of vector arises is in the development of the differential rate of strain.  Fortunately, this is also an example that \textsc{Gibbs} consider explicitly in his writing \citep[][\textsection 75, p. 162]{gibbs1901vector}.  Outlining Gibbs' development helps to establish by concrete example how Gibbs intended to express the gradient of a vector field as a dyadic.

Gibbs writes the differential of ${\bf v}$ by

\begin{align}
    d{\bf v}={\bf v}(x+dx, y+dy, z+dz) - {\bf v}(x,y,z)
\end{align}
letting ${\bf v} = v_1 {\bf i}+v_2 {\bf j} + v_3 {\bf k}$, he writes

\begin{align}
    d{\bf v} = dv_1 \, {\bf i} + dv_2 \, {\bf j} + dv_3 \, {\bf k}
\end{align}
Then, using the chain rule for differentiation in multiple dimensions, Gibbs writes

\begin{align}
    dv_1 &= d{\bf r} \cdot (\nabla v_1) \\
    dv_2 &= d{\bf r} \cdot (\nabla v_2) \\
    dv_3 &= d{\bf r} \cdot (\nabla v_3)
\end{align}
Here \citep[][\textsection 60, p 131]{gibbs1901vector},
\begin{align}
    {\bf r} &= x {\bf i} + y {\bf j} + z {\bf k} \\
   d{\bf r} &= \left( d{\bf r}\cdot {\bf i}\right)
   {\bf i} + 
   \left( d{\bf r}\cdot {\bf j} \right)
   {\bf j} + 
   \left( d{\bf r}\cdot {\bf k}\right) 
   {\bf k}
    \end{align}
Combining the steps above, the conclusion reached is the following

\begin{align}
     d{\bf v} &= d{\bf r} \cdot \left[ (\nabla v_1) \otimes {\bf i} + (\nabla v_2) \otimes {\bf j} + (\nabla v_3) \otimes {\bf k} \right]\\
\end{align}
Now,  by the definitions associated with $(\nabla v_1)\otimes {\bf i}$, etc. (cf., Eq.~\eqref{dyadform1}), we have

\begin{align}
     d{\bf v} &= d{\bf r} \cdot 
     \begin{bmatrix}
    \overset{~}{ \underset{~}{\dfrac{\partial v_1}{\partial x}}} & ~0 & 0~\\
      \underset{~}{\dfrac{\partial v_1}{\partial y}} & ~0 & 0~\\
       \underset{~}{\dfrac{\partial v_1}{\partial z} }& ~0 & 0~
     \end{bmatrix}
   +
     \begin{bmatrix}
    ~0& \overset{~}{ \underset{~}{\dfrac{\partial v_2}{\partial x}}} & ~0~  \\
      ~0& \underset{~}{\dfrac{\partial v_2}{\partial y}} & ~0~ \\
      ~0& \underset{~}{\dfrac{\partial v_2}{\partial z} }& ~0~ 
     \end{bmatrix}
     +
     \begin{bmatrix}
    ~0& ~0~ & \overset{~}{ \underset{~}{\dfrac{\partial v_3}{\partial x}}} ~\\
     ~0& ~0~ & \underset{~}{\dfrac{\partial v_3}{\partial y}} ~ \\
   ~0& ~0~ & \underset{~}{\dfrac{\partial v_3}{\partial z} }~
      \end{bmatrix}\\
      \intertext{Or,}
     d{\bf v}&=d{\bf r}\cdot
           \begin{bmatrix}
    ~\overset{~}{ \underset{~}{\dfrac{\partial v_1}{\partial x}}} & ~\overset{~}{ \underset{~}{\dfrac{\partial v_2}{\partial x}}}~ & \overset{~}{ \underset{~}{\dfrac{\partial v_3}{\partial x}}} ~\\
     ~\underset{~}{\dfrac{\partial v_1}{\partial y}}& ~\underset{~}{\dfrac{\partial v_2}{\partial y}}~ & \underset{~}{\dfrac{\partial v_3}{\partial y}} ~ \\
   ~\underset{~}{\dfrac{\partial v_1}{\partial z}}& ~\underset{~}{\dfrac{\partial v_2}{\partial z}} ~ & \underset{~}{\dfrac{\partial v_3}{\partial z} }~
      \end{bmatrix}\\
      d{\bf v}&= d{\bf r} \cdot \nabla \otimes {\bf v} 
\end{align}
Thus, one finds

\begin{equation}
        \nabla \otimes {\bf v}   =
        \begin{bmatrix}
    ~\overset{~}{ \underset{~}{\dfrac{\partial v_1}{\partial x}}} & ~\overset{~}{ \underset{~}{\dfrac{\partial v_2}{\partial x}}}~ & \overset{~}{ \underset{~}{\dfrac{\partial v_3}{\partial x}}} ~\\
     ~\underset{~}{\dfrac{\partial v_1}{\partial y}}& ~\underset{~}{\dfrac{\partial v_2}{\partial y}}~ & \underset{~}{\dfrac{\partial v_3}{\partial y}} ~ \\
   ~\underset{~}{\dfrac{\partial v_1}{\partial z}}& ~\underset{~}{\dfrac{\partial v_2}{\partial z}} ~ & \underset{~}{\dfrac{\partial v_3}{\partial z} }~
      \end{bmatrix}\label{gradvector}
\end{equation}\\
There is no danger of mistaking Gibbs' intent here.  The dyad represented by the symbols $\nabla \otimes {\bf v}$ in Gibbsian vector analysis can  be interpreted \emph{only} as given in Eq.~\eqref{gradvector}.  
Implicit in this analysis of the differential rate of strain, however, is the fact that the differential $d{\bf r}$ is represented by a right-hand multiplication with the tensor (or, the dyadic is in postfactor position).  Alternatively, one could have taken the transpose (complement) of $\nabla \otimes {\bf v}$ 
to express the result in \emph{prefactor} form

\begin{equation}
     d{\bf v}=  (\nabla \otimes {\bf v})^\dagger\cdot d{\bf r}
\end{equation}
leading to the result

\begin{equation}
        (\nabla \otimes {\bf v})^\dagger   =
        \begin{bmatrix}
    ~\overset{~}{ \underset{~}{\dfrac{\partial v_1}{\partial x}}} & ~\overset{~}{ \underset{~}{\dfrac{\partial v_2}{\partial y}}}~ & \overset{~}{ \underset{~}{\dfrac{\partial v_3}{\partial z}}} ~\\
     ~\underset{~}{\dfrac{\partial v_1}{\partial x}}& ~\underset{~}{\dfrac{\partial v_2}{\partial y}}~ & \underset{~}{\dfrac{\partial v_3}{\partial z}} ~ \\
   ~\underset{~}{\dfrac{\partial v_1}{\partial x}}& ~\underset{~}{\dfrac{\partial v_2}{\partial y}} ~ & \underset{~}{\dfrac{\partial v_3}{\partial z} }~
      \end{bmatrix}\label{gradvectorT}
\end{equation}\\
It is the form given by Eq.~\eqref{gradvectorT} that is equivalent to the \emph{alternative form} of the gradient of a vector field stated by Eq.~\eqref{altform}.  

To be clear on this point, many authors use the alternate form as the definition of the dyadic representing the gradient of a vector field

\begin{equation}
        (\nabla \otimes {\bf v})^*   =
        \begin{bmatrix}
    ~\overset{~}{ \underset{~}{\dfrac{\partial v_1}{\partial x}}} & ~\overset{~}{ \underset{~}{\dfrac{\partial v_2}{\partial y}}}~ & \overset{~}{ \underset{~}{\dfrac{\partial v_3}{\partial z}}} ~\\
     ~\underset{~}{\dfrac{\partial v_1}{\partial x}}& ~\underset{~}{\dfrac{\partial v_2}{\partial y}}~ & \underset{~}{\dfrac{\partial v_3}{\partial z}} ~ \\
   ~\underset{~}{\dfrac{\partial v_1}{\partial x}}& ~\underset{~}{\dfrac{\partial v_2}{\partial y}} ~ & \underset{~}{\dfrac{\partial v_3}{\partial z} }~
      \end{bmatrix}\label{gradvectorT2}\\
\end{equation}
While a careful presentation using the alternative form given by Eq.~\eqref{gradvectorT2} can lead to tensorially correct results, the distinction among the presentations is important. Two things must be recognized.

\begin{enumerate}
    \item The gradient of a vector field in the form of Eq.~\eqref{gradvectorT2} is \emph{not} a part of the Gibbsian formalism of vector analysis. 
    
    \item The presentation of the \emph{alternative form} of Eq.~\eqref{gradvectorT2} as defining the gradient of a vector field creates substantial opportunity for confusion; the dyadic thus formed is done in a manner where the roles of the antecedent and the consequent are reversed.  Thus, such a form does not immediately follow from the basic rules of constructing dyads when the symbol $\nabla$ is treated as a vector (or more properly, a vector operator).
\end{enumerate}

\subsection{Clarifying the difference in definitions: Decomposition of the differential strain}

It is clear that the two competing definitions of the gradient of a vector field create no problems under the circumstances that the gradient field is a symmetric tensor (in general, it will not be), or if there is a later symmetrizing operation conducted on the tensor.  

It is not difficult to find an example where the difference between the two forms of the gradient of a vector field can be seen to make a \emph{substantive} difference in the results; the conventional decomposition of the stress tensor in fluid (or solid) mechanics provides an interesting example that illustrates the issue. 

\subsubsection{Gibbsian form}

In the analysis of stresses in fluid mechanics, it is frequent to adopt the Stokes-Gibbs decomposition theorem [\citealp[][]{stokes1845}; \citealp[][Chp.~III, \textsection 137]{gibbs1884vector}]
\begin{align}
    d{\bf v}~~&= d{\bf r} \cdot \dfrac{1}{2}\left\{ 
    \left[ \nabla \otimes {\bf v}+ (\nabla \otimes {\bf v}^\dagger) \right]
    +\left[ \nabla \otimes {\bf v}- (\nabla \otimes {\bf v}^\dagger) \right]
    \right\} \nonumber\\
    &=d{\bf r}\cdot({\mathbfsf{d}}+\boldsymbol{\Omega}) \label{decomp}
    \intertext{where we note that the differential velocity is thus decomposed in \emph{postfactor} form by}
     d{\bf v}&= \underbrace{ ~d{\bf r}\cdot {\mathbfsf{d}}~}_{\substack{\textrm{deformation}}}
     + ~~\underbrace{~d{\bf r}\cdot\boldsymbol{\Omega}~}_{\textrm{rotation}}
\end{align}
Note that $\mathbfsf{d}$ is symmetric, so $\mathbfsf{d}=\mathbfsf{d}^\dagger$ are identical tensors.  Here, $\mathbfsf{d}$ is frequently called the \emph{rate of strain} or \emph{rate of deformation} tensor \citep[][p.~46]{truesdell1954} in fluid mechanics.  The dyad $\boldsymbol{\Omega}$ is referred to as the \emph{rate of rotation tensor}, \emph{vorticity tensor} \citep{serrin1959}, or \emph{spin tensor} \citep[][\textsection 86]{truesdell1960classical}; it is antisymmetric.  by comparison with Eq.~\eqref{decomp}, the vorticity tensor is

\begin{equation}
   \boldsymbol{\Omega}= \left[ \nabla \otimes {\bf v}- 
   {\left(\nabla \otimes {\bf v}\right)}^\dagger \right]
\end{equation}
or, explicitly listing the components of this dyadic

\begin{equation}
         \boldsymbol{\Omega}=
           \begin{bmatrix}
    ~\overset{~}{ \underset{~}{0}} & ~\overset{~}{ \underset{~}{\dfrac{1}{2}
    \left( \dfrac{\partial v_2}{\partial x} -\dfrac{\partial v_1}{\partial y} \right)} }~ 
    & \overset{~}{ \underset{~}{-\dfrac{1}{2}\left( \dfrac{\partial v_1}{\partial z} -\dfrac{\partial v_3}{\partial x} \right)}} ~\\
      ~\overset{~}{ \underset{~}{-\dfrac{1}{2}
    \left( \dfrac{\partial v_2}{\partial x} -\dfrac{\partial v_1}{\partial y} \right)} }~ 
    & ~\overset{~}{ \underset{~}{0}} 
    &\overset{~}{ \underset{~}{\dfrac{1}{2}\left( \dfrac{\partial v_3}{\partial y} -\dfrac{\partial v_2}{\partial z} \right)}} ~\\
     ~\overset{~}{ \underset{~}{\dfrac{1}{2}
    \left( \dfrac{\partial v_1}{\partial z} -\dfrac{\partial v_3}{\partial x} \right)} }~ 
    &\overset{~}{ \underset{~}{-\dfrac{1}{2}\left( \dfrac{\partial v_3}{\partial y} -\dfrac{\partial v_2}{\partial z} \right)}} ~
    & 0~
      \end{bmatrix}
      \label{rot1}
\end{equation}
to be used as a postfactor (left matrix multiplication)

(e.g., as used by \citet[][p. 58]{truesdell1954} and by \citet{serrin1959}).  this  is often the  definition provided for the vorticity; the expression is in Gibbsian form, and it is correct only when $\boldsymbol{\Omega}$ is used in \emph{postfactor} position.

\subsubsection{Alternative form}

In a number of texts, the vorticity tensor is defined by

\begin{equation}
   (\boldsymbol{\Omega})^*= \left[ (\nabla\otimes {\bf v})^* - \left( (\nabla \otimes {\bf v})^*\right)^\dagger \right]
\end{equation}
where $(\nabla \otimes {\bf v})^*$ is defined by Eq.~\eqref{altform} (e.g., 
Eq.~\eqref{rotwrong} (e.g., \citealp[][p.~89]{aris1962vectors}; \citealp[][p.~152]{whitaker1968}; \citealp[][p.~32]{slattery1972}; \citealp[][p.~23]{pope2000}).  While this result does generate the correct \emph{prefactor} form for the rate of rotation tensor, it uses a definition for the gradient of a vector field that is the transpose of that established by Gibbs.  This  leads to the  componentwise  result

\begin{equation}
      (\boldsymbol{\Omega})^*  = \boldsymbol{\Omega}^\dagger=
           \begin{bmatrix}
    ~\overset{~}{ \underset{~}{0}} 
    & ~\overset{~}{ \underset{~}{-\dfrac{1}{2}
    \left( \dfrac{\partial v_2}{\partial x} -\dfrac{\partial v_1}{\partial y} \right)} }~ 
    & \overset{~}{ \underset{~}{\dfrac{1}{2}\left( \dfrac{\partial v_1}{\partial z} -\dfrac{\partial v_3}{\partial x} \right)}} ~\\
      ~\overset{~}{ \underset{~}{\dfrac{1}{2}
    \left( \dfrac{\partial v_2}{\partial x} -\dfrac{\partial v_1}{\partial y} \right)} }~ 
    & ~\overset{~}{ \underset{~}{0}} 
    &\overset{~}{ \underset{~}{-\dfrac{1}{2}\left( \dfrac{\partial v_3}{\partial y} -\dfrac{\partial v_2}{\partial z} \right)}} ~\\
     ~\overset{~}{ \underset{~}{-\dfrac{1}{2}
    \left( \dfrac{\partial v_1}{\partial z} -\dfrac{\partial v_3}{\partial x} \right)} }~ 
    &\overset{~}{ \underset{~}{\dfrac{1}{2}\left( \dfrac{\partial v_3}{\partial y} -\dfrac{\partial v_2}{\partial z} \right)}} ~
    & 0~
      \end{bmatrix}
      \label{rotwrong}
\end{equation}\\
and can only to be used properly as a prefactor (right matrix multiplication).  Again, we note that the ``$(\boldsymbol{\Omega})^*$" notation is not usually used in typesetting, and we have adopted it to here distinguish between the two forms for the rate of rotation tensor.
Under these circumstances, the differential velocity would be decomposed in \emph{prefactor} form by

\begin{equation}
    d{\bf v}= \underbrace{ ~ {\mathbfsf{d}}\cdot d{\bf r}~}_{\substack{\textrm{deformation}}}
     + ~~\underbrace{~\boldsymbol{\Omega}^\dagger\cdot d{\bf r}~}_{\textrm{rotation}}
\end{equation}
where here we have used the fact that $\mathbfsf{d}=\mathbfsf{d}^\dagger$.  Thus, for infinitesimal displacements of a material body, when expressed in a non-Gibbsian prefactor form, the rotation tensor must be taken as the transpose of the Gibbsian definition expressed by $\boldsymbol{\Omega}$ if one is to maintain the proper direction of rotation.

It is likely the confusion surrounding the gradient of a vector has arisen, at least in part, because it is often only the symmetric component of the dyadic (e.g., the component $\mathbfsf{d}$ in the discussion above) that appears in many physical applications.  For example, \emph{only the symmetric component} $\mathbfsf{d}$ appears in the Navier-Stokes equations for an incompressible Newtonian fluid (although no such simplification will generally be true for non-Newtonian fluids).  However, when one examines the kinematics of flow, both $\mathbfsf{d}$ and $\boldsymbol{\Omega}$ are required.  Under such conditions,  the definition of the gradient of the velocity field needs careful consideration if one is proposing to be following the intent of Gibbs and the notations that are defined by him.

\section{Conclusions and Perspective}

Gibbs notation for vectors is one of the most widely adopted methods for presenting complex equations in mathematics, physics, and engineering.  The dyadic is a lesser-used construct, although one that Gibbs thought of as a significant achivement (as evidened by his statement  ``Let us now return to the incleterminate product, which I am
inclined to regard as the most important of all \ldots'' during his 1886 address \citep[][p.~24]{gibbs1886multiple}).  The gradient of a vector, presented as a dyadic, has unforntunately been presented in two formats, only one of which is consistent in notation with Gibbs' dyadics.   A careful assessment Gibbs' representation for dyads of the form $\nabla\otimes {\bf v}$ results in the following facts that can be made about such dyads.

\begin{enumerate}
    \item The dyad $\nabla\otimes {\bf v}$ is given in tensor form by 
    \begin{equation}
        \nabla\otimes {\bf v} \equiv \frac{\partial v_j}{\partial x_i}{\bf e}_i\otimes{\bf e}_j
        \label{graddef}
    \end{equation}
    and this representation is the only form that corresponds with Gibbs' notation. \\
    
    \item If ${\bf v}$ is a velocity vector, and $d{\bf r}$ a differential displacement, then the differential rate of strain can be represented by
    \begin{equation}
       d{\bf v} =  d{\bf r}\cdot \nabla \otimes {\bf v} 
    \end{equation}
    in postfactor form or 
        \begin{equation}
       d{\bf v} =   (\nabla \otimes {\bf v})^\dagger \cdot d{\bf r}
    \end{equation}
    in prefactor form.  Regardless of the form used to express the result, the dyad $\nabla\otimes {\bf v}$ is defined in the Gibbsian sense as in Eq.~\eqref{graddef}.\\
    
    \item The solid body rotation tensor defined in postfactor form is given by
    
    \begin{equation}
       d{\bf r}\cdot \boldsymbol{\Omega} = d{\bf r}\cdot\left[ \nabla \otimes {\bf v}- 
          {\left(\nabla \otimes {\bf v}\right)}^\dagger \right]
    \end{equation}
    and, in prefactor form by
        \begin{equation}
           \boldsymbol{\Omega}^\dagger\cdot d{\bf r} = \left[ \nabla \otimes {\bf v}- 
              {\left(\nabla \otimes {\bf v}\right)}^\dagger \right]^\dagger \cdot d{\bf r}
    \end{equation}
     As is true for conclusion 2, the dyad $\nabla\otimes {\bf v}$ is defined in the Gibbsian sense as in Eq.~\eqref{graddef}.\\
    
    \item Frequently, the vorticity matrix ($\boldsymbol{\Omega}$ in the developments above) is specified without a clear indication as to whether the representation is in prefactor or postfactor form.  This leads to an indeterminate presentation.  For clarity, it is may be useful to express how subsequent multiplications are intended (e.g., as a postfactor for $\boldsymbol{\Omega}$, or as a prefactor for $\boldsymbol{\Omega}^\dagger$).   \\
    
    \item Because there of the past ambiguity in the use of the symbol $\nabla {\bf v}$, it is suggested that when one intends to use Gibbsian notation for the dyadic, that this be emphasized by explicit use of the $\otimes$ symbol, especially when the gradient of a vector field is represented.  Thus, $\nabla \otimes {\bf v}$ would be used for a dyad in postfactor position, whereas  $(\nabla \otimes {\bf v})^\dagger$ would be used for a dyad in prefactor position.  \\
    
    \item The modern tools of geometric algebra can be adapted to represent the results that we have discussed above.  It is especially notable that these results are \emph{consistent} with the form defining the gradient of a vector field in Gibbs' notation.  These results are discussed further in Appendix \ref{appendixB}.

\end{enumerate}

\bmhead{Data Availability Statement}
Data sharing not applicable to this article as no datasets were generated or analysed during the current study.

\bmhead{Acknowledgments}
This material is based upon work supported by the U.S. Department of Energy, Office of Science, Office of  Basic Energy Sciences (Geosciences) under Award Number DE‐SC0021626.

\bibliography{gradvec}

\begin{appendices}

\section{Nonion Forms}\label{appendixA}

Gibbs identified a basis set for second-order tensors formed by the dyads of the unit vectors.  In general, Gibbs considered non-orthogonal basis vectors, although in much of his work a Cartesian coordinate system is implied.  For a right-handed Cartesian coordinate systems, \citet{gibbs1884vector} denoted the unit vectors ${\bf i}$, ${\bf j}$, and ${\bf k}$ by  ${\bf i}={\bf e}_1=(1,0,0)$, ${\bf j}={\bf e}_2=(0,1,0)$, and ${\bf k}={\bf e}_3=(0,0,1)$.  Here, the correspondence with the more modern numerically-indexed basis vector notation ${\bf e}_i$ is established for reference. The dyads formed by the nine possible combinations of the unit vectors form a basis for all second-order tensors, and this is particularly obvious for Cartesian coordinate systems.  Following the convention of the main body of the text, dyadics are explicitly denoted by the $\otimes$ symbol.  The basis dyadics are then given by 

\begin{equation}
    {\bf i}\otimes {\bf i} ={\bf e}_1\otimes{\bf e}_1=
    \begin{bmatrix}
    1 & 0 & 0\\
    0 & 0 & 0\\
    0 & 0 & 0
    \end{bmatrix}
    ,~~
        {\bf i}\otimes {\bf j} ={\bf e}_1\otimes{\bf e}_2=
    \begin{bmatrix}
    0 & 1 & 0\\
    0 & 0 & 0\\
    0 & 0 & 0
    \end{bmatrix}
    ,
    ~\textrm{etc.}
\end{equation}
From these definitions, any second-order tensor 

\begin{equation}
{\mathbfsf T} = 
\begin{bmatrix}
T_{11} & T_{12} & T_{13} \\
T_{21} & T_{22} & T_{23} \\
T_{31} & T_{32} & T_{33} 
\end{bmatrix}
\end{equation}
can be written in terms of the Cartesian basis vectors by

\begin{align}
{\mathbfsf T} &= T_{11}{\bf e}_1\otimes{\bf e}_1 + T_{12}{\bf e}_1\otimes{\bf e}_2+T_{13}{\bf e}_1\otimes{\bf e}_3 \nonumber\\
&+ T_{21}{\bf e}_2\otimes{\bf e}_1 + T_{22}{\bf e}_2\otimes{\bf e}_2+T_{23}{\bf e}_2\otimes{\bf e}_3\nonumber\\
&+T_{31}{\bf e}_3\otimes{\bf e}_1 + T_{32}{\bf e}_3\otimes{\bf e}_2+T_{33}{\bf e}_3\otimes{\bf e}_3
\end{align}
\citet[][\textsection 101]{gibbs1884vector} referred to this as the \emph{nonion} form of a dyad.

\section{Representation of Dyadics via Geometric Algebra}\label{appendixB}

Much has been made of the competition of Gibbs' vector analysis and Hamilton's quaternions for describing mechanics in three-dimensional space.  We will not attempt to address this (fortunately short in duration) division between the competing vector schemes in any detail; for interested readers, the details of this moderately untoward episode are presented in \citet[][Chp.~Six]{crowe1967history}.  At the center of this controversy was the topic of which form of \emph{multiple algebras} (as they were called at the time) were best suited as the language of physics.  Multiple algebras were mathematical systems that allowed the combination of elements with different dimension (e.g., scalar and vector quantities could appear as a sum in a single term), in a consistent algebraic context; these were developed by originally by H. Grassman \citep{grassmann1_1844,grassmann2_1862ausdehnungslehre}, though they remained somewhat obscure until the the 1880s, when they were discovered by the main stream researchers working on vector theory.  Hamilton's quaternions were an example of a multiple algebra, but one that would be considered a subset of Grassman's more comprehensive framework.  The Gibbsian vector algebra could also be considered as an even more restricted subset of the Grassman framework.  

While it is often supposed that Gibbs himself was not supportive of quaternions, there is much evidence to the contrary.  Gibbs himself provided an account of the two-dimensional analogue to quaternions (i.e., vectors expressed in the complex plane, which he calls \emph{bivectors}, \citep[][``Note on Bivector Analysis'', p.~76 ]{gibbs1884vector}.  Even when his version of vector algebra was unkindly (and incorrectly) attacked by those supporting the quaternion perspective (again, see \citet[][Chp.~Six]{crowe1967history} for the less than seemly details), he was not baited; his responses focused on solid and unimpassioned refutations of the various (and largely spurrious) critiques of his work.   In short, Gibbs was supportive of the use of multiple algebras, including quaternions.  He did, however, assign priority to Grassman rather than Hamilton.  He also did not see a need to choose one framework (quaternions) over another (vector analysis), noting that they each had value in particular applications.

Gibbs' was not one known to public discussion of conjecture, preferring to restrict himself to what could be more solidly proven.  One exception, however, is is thoughts about the future and usefulness of multiple algebras.  Gibbs' own work on bivectors was an acknowledgement of the value of multiple algebras, and he was an enthusiastic promoter of Grassman, taking some pains to assure that his work became known. In Gibbs' 1886 address to the American Association for the Advancement of Science (Mathematics and Astronomy Section) he made the explicit connection between Grassman's algebra and the work of Hamilton \citep{gibbs1886multiple}, making clear his respect for the quaternion perspective of Hamilton.

\begin{quote}
    The failure of Mo\"bius, Hamilton, Grassmann, Saint-Venant to make an immediate impression upon the course of mathematical thought in any way commensurate with the importance of their discoveries is the most conspicuous evidence that the times were not ripe for the methods which they sought to introduce.  A satisfactory theory of the  imaginary quantities of ordinary algebra, which is essentially a simple case of multiple algebra, with difficulty obtained recognition in the first third of this century. We must observe that this \emph{double algebra}, as it has been called, was not sought for or invented; it forced itself, unbidden, upon the attention of mathematicians, and with its rules already formed.\ldots  
    
    The application of double algebra to the geometry of the plane suggested not unnaturally to Hamilton the idea of a triple algebra  which should be capable of a similar application to the geometry of three dimensions. He \ldots  ~discovered at length a quadruple algebra, \emph{quaternions}, which answered his purpose, thus satisfying, as he says in one of his letters, an intellectual want which had haunted him at least fifteen years.
\end{quote}
The remainder of the address focuses on the history and potential utility of multiple algebras.  With this expressed interest and acknowledgement of Hamilton's contributions to the field, it is hard to suggest that Gibbs was other than supportive of the concept underlying multiple algebras, which included quaternions.

There is one additional consideration that should be noted.  While Gibbs was a supporter of multiple algebras in principle, he did not publish extensively on the topic himself. While Gibbs personally promoted the ideas of Grassmann and, subsequently,  Clifford \citet{clifford1878}] (who he mentions in the introduction to his \emph{Elements of Vector Analysis}), the history presented by \citet{crowe1967history} argues that his knowledge of their work was most likely superficial.  This perspective, however, has been at least partially challenged by one of the last publications of Gibbs' student and academic heir, E.B. Wilson. In 1961, \citet{wilson1961last}  summarized some of the unpublished notes of Gibbs, indicating that his personal research on multiple algebras went much deeper than was perhaps previously recognized .  

Although we can only speculate, one of the problems may have been that dyadics, which were one of the central developments of Gibbs' vector algebra, had no obvious extension to multiple algebras.  To address this issue directly, in Appendix \ref{appendixB} we have sought to re-express the results of Gibbs on the differential strain in terms of Geometric Algebra (which arose from both Grassman's and, especially, Clifford's work \citep{clifford1878}) in the modern version of multiple algebra, known as geometric algebra \citep{hestenes2012clifford}.
There have been continuing efforts to connect geometric algebra with Gibbs' dyads (cf., \citet{lindell2000}), and this appendix continues in that direction.

\subsection{Geometric Algebra Fundamentals.}

Geometric algebras are real Clifford algebras augmented with some additional operator notation.  We use the geometric algebra for $\mathbb{R}^{3}$ to
construct multivector representations of the differential strain rate, and it's symmetric and antisymmetric tensor product components.  Questions of which tensor product representation to use, and how to interpret dyadic quantities, are eliminated by the intrinsic vector multiplication operation product provided by geometric algebra, suggesting possible value of future research in this direction.

A geometric algebra \(G(V)\), is a vector space generated from a dot product space \(V\),
the elements of which are multivectors (sums of scalars, vectors, and products of vectors), where vectors \(\mathbf{v} \in V\) are subject to a multiplication rule \(\mathbf{v}^2 = \mathbf{v} \cdot \mathbf{v}\).
In the interest of brievity, the reader is referred to other sources, such as
\citep{doran2003gap},
\citep{dorst2007gac},
\citep{aMacdonaldLAGC}, and
\citep{pjootGAEE} for a more thorough grounding in geometric algebra fundamentals.
In this appendix we restrict attention to the geometric algebra generated by \(V = \mathbb{R}^3\), and present a
minimal set of identities, definitions and properties.

Products in geometric algebras are not generally commutative.
In particular, products of parallel vectors commute, and it can be shown that products of perpendicular vectors anticommute, for example \(\mathbf{e}_1 \mathbf{e}_2 = -\mathbf{e}_2 \mathbf{e}_1\).
An implication of such anticommutivity is that Eq.~\eqref{altform} is incorrect in geometric algebra.

A product of \(k\) orthogonal vectors is called a blade, where \(k\) is called the grade, and
where 0-blades and 1-blades are defined respectively as scalars and vectors.
Respective examples of 0,1,2,3 blades are \(1, \mathbf{e}_3, (\mathbf{e}_1 + \mathbf{e}_3)\mathbf{e}_2, \mathbf{e}_1 \mathbf{e}_2 \mathbf{e}_3\).
The geometric algebra for higher dimensional spaces such as $\mathbb{R}^{4}$ may have elements like \( \mathbf{e}_1 \mathbf{e}_2 + \mathbf{e}_3 \mathbf{e}_4 \) that cannot be represented as a product of two vectors.  Such general elements are called k-vectors, where 0-vectors,1-vectors,2-vectors,3-vectors are respectively also referred to as scalars, vectors, bivectors, trivectors.
However, for $\mathbb{R}^3$, any element of a given grade can be represented as a blade.

A k-blade is irreducible, whereas a general multivector can always be reduced to a sum of k-blades.  For example, one can show that \( \mathbf{e}_1(\mathbf{e}_2 + \mathbf{e}_3) \mathbf{e}_1\mathbf{e}_2 = -1 - \mathbf{e}_2 \mathbf{e}_3 \), a sum of a 0-blade (scalar) and a 2-blade (bivector), with respective grades 0,2.

The fundamental operator in geometric algebra is that of grade selection, designated with angle brackets and a numeric suffix.  The grade-k selection of a multivector \( M \) is designated \( {\left\langle M \right\rangle}_{k} \).
For example, given \( M = 1 + \mathbf{e}_3 + (\mathbf{e}_1 + \mathbf{e}_3)\mathbf{e}_2 + \mathbf{e}_1 \mathbf{e}_2 \mathbf{e}_3 \)
\begin{equation}\label{eqn:gradvec:40}
\begin{aligned}
   {\left\langle M \right\rangle}_{0} 
   &= 1 \\
   {\left\langle M \right\rangle}_{1}
   &= \mathbf{e}_3 \\
   {\left\langle M \right\rangle}_{2}
   &= \mathbf{e}_1 \mathbf{e}_2 - \mathbf{e}_2 \mathbf{e}_3 \\
   {\left\langle M \right\rangle}_{3}
   &= \mathbf{e}_1 \mathbf{e}_2 \mathbf{e}_3.
\end{aligned}
\end{equation}
Specific grade selection operations have enough utility to justify specialized operator notation.  In particular, given a k-blade \(B_k\) and a j-blade \(B_j\) we may define
\begin{equation}\label{eqn:gradvec:60}
\begin{aligned}
   B_k \cdot B_j &= {\left\langle B_k B_j \right\rangle}_{\lvert k - j \rvert} \\
   B_k \wedge B_j &= {\left\langle B_k B_j \right\rangle}_{k + j}.
\end{aligned}
\end{equation}
Given a vector \( \mathbf{a} \), it is possible to show that
\begin{equation}\label{eqn:gradvec:80}
\begin{aligned}
   B_k \cdot \mathbf{a} &= \frac{1}{2} \left( B_k \mathbf{a} + (-1)^{k+1} \mathbf{a} B_k \right) \\
   B_k \wedge \mathbf{a} &= \frac{1}{2} \left( B_k \mathbf{a} - (-1)^{k+1} \mathbf{a} B_k \right).
\end{aligned}
\end{equation}
Special cases of dot and wedge product operations used in this appendix include
\begin{equation}\label{eqn:gradvec:100}
\begin{aligned}
\mathbf{a} \cdot \mathbf{b}  &= \frac{1}{2} 
   \left( \mathbf{a} \mathbf{b} + \mathbf{b} \mathbf{a} \right)
   = \mathbf{b} \cdot \mathbf{a} \\
\mathbf{a} \wedge \mathbf{b} &= \frac{1}{2} 
   \left( \mathbf{a} \mathbf{b} - \mathbf{b} \mathbf{a} \right)
   = -\mathbf{b} \wedge \mathbf{a} \\
B \cdot \mathbf{a}    &= \frac{1}{2} 
   \left( B \mathbf{a} - \mathbf{a} B \right)
   = - \mathbf{a} \cdot B,
\end{aligned}
\end{equation}
where \( \mathbf{a}, \mathbf{b} \) are vectors, and \(B\) is a 2-blade.

We will also make use of the fundamental identity
\begin{equation}\label{eqn:gradvec:140}
\begin{aligned}
   \mathbf{a} \mathbf{b} = \mathbf{a} \cdot \mathbf{b} + \mathbf{a} \wedge \mathbf{b},
\end{aligned}
\end{equation}
and the distribution identity
\begin{equation}\label{eqn:gradvec:120}
\begin{aligned}
   \mathbf{a} \cdot 
   \left( \mathbf{b} \wedge \mathbf{c} \right)
   = 
   \left( \mathbf{a} \cdot \mathbf{b} \right)
   \mathbf{c} - 
   \left( \mathbf{a} \cdot \mathbf{c} \right)
   \mathbf{b}.
\end{aligned}
\end{equation}
\subsection{The Differential Rate of Strain in Geometric Algebra}
The differential rate of strain need may be expressed in scalar operator form as
\begin{equation}\label{eqn:gradvec:160}
   d\mathbf{v} = \frac{\partial \mathbf{v}}{x_i} dx_i = \left( d\mathbf{x} \cdot \nabla \right) \mathbf{v}.
\end{equation}
A coordinate expansion yields
\begin{equation}\label{eqn:gradvec:180}
   d\mathbf{v} = dx_i \partial_i v_j \mathbf{e}_j,
\end{equation}
where we introduced shorthand \( \partial_i = \partial/{\partial x_i} \).  Decomposition into
symmetric and antisymmetric tensor components yields
\begin{equation}\label{eqn:gradvec:200}
d\mathbf{v} =
\frac{1}{2} dx_i \left( \partial_i v_j + \partial_j v_i \right) 
\mathbf{e}_j +
\frac{1}{2} dx_i \left( \partial_i v_j - \partial_j v_i \right) 
\mathbf{e}_j.
\end{equation}
The first term is the contribution of the symmetric component of \( \nabla \otimes \mathbf{v} \), whereas the second is the contribution of the antisymmetric component.
In vector notation, these are respectively
\begin{equation}\label{eqn:gradvec:220}
\begin{aligned}
   d\mathbf{x} \cdot \mathbf{d} 
   &= \frac{1}{2} \left( (d\mathbf{x} \cdot \nabla) \mathbf{v} \right) + \nabla \left( d\mathbf{x} \cdot \mathbf{v} \right) \\
   d\mathbf{x} \cdot \boldsymbol{\Omega} 
   &= \frac{1}{2} \left( (d\mathbf{x} \cdot \nabla) \mathbf{v} - \nabla \left( d\mathbf{x} \cdot \mathbf{v} \right) \right).
\end{aligned}
\end{equation}
The antisymmetric term \( \boldsymbol{\Omega} \), using Eq.~\eqref{eqn:gradvec:120}, can be put into geometric algebra form as
\begin{equation}\label{eqn:gradvec:240}
d\mathbf{x} \cdot \boldsymbol{\Omega} = \frac{1}{2} d\mathbf{x} \cdot \left( \nabla \wedge \mathbf{v} \right),
\end{equation}
allowing the compact identification
\begin{equation}\label{eqn:gradvec:260}
\boldsymbol{\Omega} = \frac{1}{2} \left( \nabla \wedge \mathbf{v} \right).
\end{equation}

The material is said to be incompressible if \( \nabla \cdot \mathbf{v} = 0 \).  Let's also observe how our representations are altered by compressibility.
We may compute the portion of the strain rate that is invariant with respect to compressibility by subtracting off a divergence term
\begin{equation}\label{eqn:gradvec:280}
\begin{aligned}
d\mathbf{v} - d\mathbf{x} \left( \nabla \cdot \mathbf{v} \right)
&=
\left( d\mathbf{x} \cdot \nabla \right) 
\mathbf{v} -
d\mathbf{x} \left( \nabla \cdot \mathbf{v} \right)
\\
&=
\nabla \cdot \left( d\mathbf{x} \wedge \mathbf{v} \right),
\end{aligned}
\end{equation}
also using the distribution identity, which splits the strain rate explicitly into compression sensitive and insensitive components
\begin{equation}\label{eqn:gradvec:300}
   d\mathbf{v} = \left( d\mathbf{x} \cdot \nabla\right) 
   \mathbf{v}
   =
   d\mathbf{x} \left( \nabla \cdot \mathbf{v} \right)
   + \nabla \cdot \left( d\mathbf{x} \wedge \mathbf{v} \right)
   .
\end{equation}

Applied to Eq.~\eqref{eqn:gradvec:220} we may compute the incompressible portions of the symmetric and antisymmetric tensors
\begin{equation}\label{eqn:gradvec:320}
\begin{aligned}
(d\mathbf{x} \cdot \nabla) \mathbf{v} + \nabla \left( d\mathbf{x} \cdot \mathbf{v} \right)
- d\mathbf{x} 
\left( \nabla \cdot \mathbf{v} \right)
&=
   \nabla \left( d\mathbf{x} \cdot \mathbf{v} \right)
   + \nabla \cdot 
   \left( d\mathbf{x} \wedge \mathbf{v} \right) \\
   &=
{\left\langle  \nabla \left( d\mathbf{x} \cdot \mathbf{v} \right) + \nabla \cdot \left( d\mathbf{x} \wedge \mathbf{v} \right)  \right\rangle}_{1} \\
&=
{\left\langle  \nabla \left( d\mathbf{x} \cdot \mathbf{v} \right) + \nabla \left( d\mathbf{x} \wedge \mathbf{v} \right)  \right\rangle}_{1} \\
&=
{\left\langle  \nabla d\mathbf{x} \mathbf{v}  \right\rangle}_{1},
\end{aligned}
\end{equation}
and similarly
\begin{equation}\label{eqn:gradvec:340}
\begin{aligned}
(d\mathbf{x} \cdot \nabla) \mathbf{v} - \nabla 
\left( d\mathbf{x} \cdot \mathbf{v} \right)
- d\mathbf{x} 
\left( \nabla \cdot \mathbf{v} \right)
&=
  -\nabla 
  \left( d\mathbf{x} \cdot \mathbf{v} \right)
  + \nabla \cdot 
  \left( d\mathbf{x} \wedge \mathbf{v} \right) \\
&=
  -\left( d\mathbf{x} \cdot \mathbf{v} \right)
  \nabla - 
  \left( d\mathbf{x} \wedge \mathbf{v} \right)
  \cdot \nabla \\
&=
  - {\left\langle  d\mathbf{x} \mathbf{v} \nabla  \right\rangle}_{1},
\end{aligned}
\end{equation}
where we allow the gradient to act bidirectionally,
but taking care not to assume commutivity.  That is, for multivectors \(M,N\), the gradient's action is \(M \nabla N = \partial_i(M \mathbf{e}_i N)\).

The split of the symmetric and antisymmetric tensor components into the divergence and non-divergence terms is therefore
\begin{equation}\label{eqn:gradvec:360}
\begin{aligned}
   d\mathbf{x} \cdot \mathbf{d} &= \frac{1}{2} 
   \left( d\mathbf{x} \left( \nabla \cdot \mathbf{v} \right) + {\left\langle  \nabla d\mathbf{x} \mathbf{v}  \right\rangle}_{1}
   \right) \\
   d\mathbf{x} \cdot \boldsymbol{\Omega} &= \frac{1}{2} 
   \left( d\mathbf{x} \left( \nabla \cdot \mathbf{v} \right) - {\left\langle  d\mathbf{x} \mathbf{v} \nabla  \right\rangle}_{1} 
   \right).
\end{aligned}
\end{equation}
\subsection{Summary of GA results}
To summarize, we found that the rate of strain splits nicely into a divergence and non-divergence term
\begin{equation}\label{eqn:gradvec:380}
   d\mathbf{v} = 
   \left( d\mathbf{x} \cdot \nabla\right) 
   \mathbf{v} = d\mathbf{x} 
   \left( \nabla \cdot \mathbf{v} \right)
   + \nabla \cdot 
   \left( d\mathbf{x} \wedge \mathbf{v} \right).
\end{equation}
We found the GA representation of the symmetric tensor component, and computed its split into divergence and non-divergence terms
\begin{equation}\label{eqn:gradvec:400}
\begin{aligned}
   d\mathbf{x} \cdot \mathbf{d} &= \frac{1}{2} 
   \left( (d\mathbf{x} \cdot \nabla) \mathbf{v} + \nabla \left( d\mathbf{x} \cdot \mathbf{v} \right) \right) \\
                  &= \frac{1}{2} 
   \left( d\mathbf{x} \left( \nabla \cdot \mathbf{v} \right) + {\left\langle  \nabla d\mathbf{x} \mathbf{v}  \right\rangle}_{1}
   \right).
\end{aligned}
\end{equation}
Finally, we found that the antisymmetric tensor component has a particularly compact GA representation, and also computed its split into divergence and non-divergence terms
\begin{equation}\label{eqn:gradvec:420}
\begin{aligned}
   d\mathbf{x} \cdot \boldsymbol{\Omega} &= d\mathbf{x} \cdot \frac{1}{2} 
   \left( \nabla \wedge \mathbf{v} \right) \\
   &= \frac{1}{2} \left( d\mathbf{x} \left( \nabla \cdot \mathbf{v} \right) - {\left\langle  d\mathbf{x} \mathbf{v} \nabla  \right\rangle}_{1} \right).
\end{aligned}
\end{equation}




\end{appendices}




\end{document}